# Could the Fundamental Laws of Nature be Inferred Logically (Mathematically) from Only a Very Few Axioms?

**Ramin Zahedi**

This article is a summary of an expanded version Ref. [1]. In Ref. [1], as a new mathematical approach to origin of the laws of nature, using a new basic algebraic axiomatic (matrix) formalism based on the ring theory and Clifford algebras (presented in Sec.2), "it is shown that certain mathematical forms of fundamental laws of nature, including laws governing the fundamental forces of nature (represented by a set of two definite classes of general covariant massive field equations, with new matrix formalisms), are derived uniquely from only a very few axioms"; where in agreement with the rational Lorentz group, it is also basically assumed that the components of relativistic energy-momentum can only take rational values. Based on the definite mathematical formalism of this axiomatic approach, along with the C, P and T symmetries (represented by the corresponding quantum matrix operators) of the fundamentally derived field equations, it is concluded that the universe could be realized solely with the (1+2) and (1+3)-dimensional space-times. On the basis of these discrete symmetries of the derived field equations, it has been also shown that only left-handed particle fields (along with their complementary right-handed fields) could be coupled to the corresponding (any) source currents. Moreover, it is shown that the (1+3)-dimensional cases of uniquely determined two classes of general covariant field equations, represent, respectively, new massive forms of the bispinor fields of spin-2, and spin-1 particles; and (1+2)-dimensional cases of these equations represent (asymptotically) new massive forms of the bispinor fields of spin-3/2 and spin-1/2 particles, respectively. As a particular result, based on the definite formulation of the derived Maxwell equations (representing by the derived bispinor fields of spin-1 particles, including Yang-Mills equations compatible with two specified forms of gauge symmetry groups), it has been concluded that magnetic monopoles could not exist in nature. Furthermore, along with the known elementary particles, eight new elementary particles, including four new charge-less right-handed spin-1/2 fermions (two leptons and two quarks), a spin-3/2 fermion, and also three new spin-1 (massive) bosons are predicted uniquely by this axiomatic approach.

## 1. Introduction

Let we start with one of the basic ontological questions that philosophy and science can ask and investigate. Why do the known fundamental forces acting on the Universe (i.e., the four basal forces that appear to cause all the movements and interactions between the existing bodies in nature which are made from a few kinds of the elementary particles), manifest in the specific "way and form" which they do permanently? In this article we present a summary of Ref. [1], where we have considered this question and tried to find a deterministic answer for it by presenting a new mathematical axiomatic approach.

Eugene Wigner's foundational paper, "On the Unreasonable Effectiveness of Mathematics in the Natural Sciences", famously observed that purely mathematical structures and axiomatic approaches often lead to deep physical insights, in turn serving as the basis of highly successful physical theories [2]. Referring to the Oxford Encyclopedia, a law of physics (or a scientific law) is: "a theoretical principle deduced from particular facts, applicable to a defined group or class of phenomena, and expressible by the statement that a particular phenomenon always occurs if certain conditions be presented [3]. In fact, laws of physics (including the fundamental laws of nature) have been typically conclusions based on the repeated experiments and observations over many years and which have become accepted universally within the scientific communities; and, in addition, one of the most essential aims of the human race has been to acknowledge (as truths) a summary description of the natural world in the form of such fundamental laws [4, 5].

In Ref. [1], in essence, we've presented a new mathematical axiomatic approach based on the ring theory (including the integral domains) and the generalized Clifford algebra. Using this approach, first we've shown that the basic algebraic properties of natural numbers, as the most logical elements which are using for representing and describing the universe's physical laws, should be axiomatized in a matrix formalism base. Then by assuming that the components of relativistic energy-momentum (D-mo-mentum) – as one of the most elemental quantities of physics – can only take the rational values, it has been shown that the fundamental relativistic field equations of physics, that are representing the laws governing the fundamental forces of nature, could be derived uniquely from only a very few axioms.

In summary, the main steps of the procedure of this axiomatic derivation approach are as follows. Firstly, it has been shown that the relativistic energy-momentum quadratic relation, necessarily, should be linearized (and simultaneously parameterized, as necessary algebraic conditions based on the above assumption for D-momentum). Consequently, corresponding to the relativistic energy-momentum quadratic (algebraic) relation, a unique set of the systems of homogeneous linear equations (which are fully compatible with the Lorentz rational symmetry group and also the Clifford algebra) have been derived for various space-time dimensions, respectively. Subsequently, by first quantization (followed by the minimal coupling procedure) of the obtained relativistic systems of linear equations, a unique set of general relativistic "massive" (tensor) field equations (including a specific form of torsion tensor) are derived. It is shown that the derived tensor field equations (that are compatible with the Clifford algebras corresponding to the certain spin groups), quantum mechanically, represent a new unique massive tensor formalism corresponding solely to the certain bispinor fields (of spin-1/2, spin-3/2, spin-1 and spin-2 elementary particles). In this regard, based on the presented derivation approach and also assuming a set of the basic discrete symmetries, it has been shown that the obtained tensor field equations which correspond uniquely to the known fundamental forces of nature including the gravitational (Einstein), electromagnetic (Maxwell) and nuclear (Yang-Mills) field equations, are expressible solely in (1+3) space-time dimensions. In addition, on the same basis, it is also concluded that the derived relativistic massive equations which present certain cases of the bispinor fields of spin-1/2 and spin-3/2 particles corresponding to the Dirac and Rarita–Schwinger equations, are expressible solely in (1+2) space-time dimensions. As a particular consequence, it is shown that a unique massive formulation of the general theory of relativity - with a specific torsion that generates the gravitational field's mass - is obtained only by first quantization (followed by the minimal coupling procedure) of a unique set of the special relativistic algebraic matrix relations. In addition, it is also shown that the Lagrangian densities specified for the obtained new massive Lorentz covariant forms of the Maxwell's (and Yang-Mills) equations and also Dirac equation, are locally gauge invariant as well–where the invariant mass of each particle field is generated by a torsion field. In this regard and in agreement with recent experimental (astronomical) data, the invariant mass of a new particle (identified as the massive gauge boson of U(1) symmetry group) has been calculated and specified as: $m_\gamma \approx 1.470696 \times 10^{-41}$ kg, where its invariant mass is generated by a coupling torsion tensor of the background space-time geometry.

Moreover, based on the definite mathematical formalism of this axiomatic approach, along with the C, P and T symmetries (represented basically by the corresponding quantum operators) of the fundamentally derived field equations, it has been concluded that the universe could be realized solely with the (1+2) and (1+3)-dimensional space-times (where this conclusion, in particular, is based on the T-symmetry of these equations). In addition, on the basis of these discrete symmetries of derived field equations, it has been also shown that only left-handed particle fields (along with their complementary right-handed fields) could be coupled to the corresponding (any) source currents. Furthermore, it has been shown that the metric of background space-time is diagonalized for the uniquely derived fermion field equations (defined and expressed solely in (1+2)-dimensional space-time), where this property generates a certain set of additional symmetries corresponding uniquely to the "SU(2)_L × U(2)_R" symmetry group for spin-1/2 fermion fields (representing "1+3" generations of four fermions, including a group of eight leptons and a group of eight quarks), and also the "SU(2)_L × U(2)_R" and SU(3) gauge symmetry groups for spin-1 boson fields coupled to the spin-1/2 fermionic source currents. Hence, along with the known elementary particles, eight new elementary particles, including four new charge-less right-handed spin-1/2 fermions (two leptons and two quarks, that could be represented by "z_e , z_n ; z_u, z_d", where two quarks "z_u, z_d" in particular are mixed solely with the triple compositions of antiquarks in anti-baryons' structures), a spin-3/2 fermion, and also three new spin-1 massive bosons (where in particular, the new boson  is complementary right-handed particle of ordinary boson), are predicted uniquely by this axiomatic approach. Furthermore, as a particular result, based on the definite and unique formulation of the derived Maxwell's equations (and also a determined new form of the Yang-Mills equations, compatible with specific gauge symmetry groups), it has been also concluded that magnetic monopoles could not exist in nature.

## 2. Summary and Results

**2-1.** The main arguments and consequences presented in Ref. [1] follow from these three new basic assumptions:

**(1)-"A new generalization of the axiom of nonzero divisors of the domain of integer elements, based on a matrix representation of the generalized Clifford algebra ; and subsequently, constructing a basic algebraic linearization theory for the set of integers ; "**

This is one of the new principal concepts which has been presented in Ref. [1], and has been formulated as follows :

"Let $F(b_1, b_2, b_3, ..., b_s)$ be a homogeneous polynomial of degree $r \geq 2$ over the domain of integers Z ($b_p \in$ Z). We assume the following new axiom as the generalization of algebraic axiom of 'nonzero divisors',

$\exists\, n,\ \exists\, A=[a_{ij}]\in Z_{n\times n},\ T=[t_i]\in Z_{n\times 1}$ :

$$[(A\times T=0,\ T\neq 0)\wedge(A^r=F(b_1,b_2,b_3,...,b_s)I_n)]\ \Leftrightarrow\ F(b_1,b_2,b_3,...,b_s)=0 \qquad \text{(A)}$$

where $a_{ij} = \sum_{p=1}^{s} H_{ijp}b_p$, $H_{ijp}$ are coefficients (independent of integer variables $b_p$), $Z_{n\times n}$ and $Z_{n\times 1}$ are respectively the set of $n\times n$ square matrices and the set of $n\times 1$ column matrices with integer components, $T$ is a non-zero parametric column matrix and $I_n$ is the identity matrix."

As a consequential remark, it is noteworthy that since in the axiomatic relation (A), $F(b_1, b_2, b_3,..., b_s) = 0$ is a homogeneous equation ($b_p \in Z$), the relation (A) in particular, is applicable to equation $F(b_1, b_2, b_3,..., b_s) = 0$ as well, for $b_p \in Q$ where $Q$ is the set of rational numbers.

**(2)-"The components of relativistic energy-momentum (D-momentum) can only take the rational values ; "**

We assume, in general, that the components of relativistic energy-momentum (D-momentum) of a particle can only take the rational values. This basic assumption is fully compatible with the rational Lorentz group which is dense in the group of Lorentz transformations. It also is in agreement with the formalism necessary for the quantum relativistic physical theories [20]. Moreover, this assumption is also clearly compatible with the quantum circumstance in which the energy and momentum of a particle are merely transferred as integer multiples of the quantum of action (Planck constant) $h$.

Using the assumptions (1) and (2), in Ref. [1] it has been concluded, straightforwardly, that the relativistic energy-momentum homogeneous (quadratic) relation should be linearized (and simultaneously parameterized) necessarily. Consequently, a unique set of the Lorentz covariant systems of linear algebraic equations (satisfying the Clifford algebras) have been obtained for various space-time dimensions, as the modified matrix forms of the relativistic energy-momentum quadratic relation. These matrix forms have been used for a unique derivation of the fundamental field equations of physics in various space-time dimensions (see the next basic assumption (3)).

**(3)-"The general covariant forms of all the fundamental field equations of physics, corresponding to the laws governing the fundamental forces of nature, are derived solely by first quantization (followed by the minimal coupling procedure) of the relativistic algebraic matrix equations obtained on the basis of assumption (2), by linearization (and si-**

**multaneously parameterization) of the relativistic energy-momentum quadratic relation."**

In addition, in Ref. [1] we assume that the source-free cases of the obtained field equations have also certain discrete symmetries separately.

Following is a summary description of some notable consequences of the axiomatic approach presented in Ref. [1], where the metric signature $(+ - \ldots -)$, the geometrized units and the following "sign" conventions have been used :

− The Riemann curvature, Ricci and Einstein tensors :

$$R^{\rho}{}_{\sigma\mu\nu} = \partial_{\nu}\Gamma^{\rho}{}_{\mu\sigma} + \Gamma^{\rho}{}_{\nu\lambda}\Gamma^{\lambda}{}_{\mu\sigma} - \partial_{\mu}\Gamma^{\rho}{}_{\nu\sigma} - \Gamma^{\rho}{}_{\mu\lambda}\Gamma^{\lambda}{}_{\nu\sigma} \,, \nabla_{\sigma}R_{\mu\nu\rho}{}^{\sigma} = \nabla_{\nu}R_{\mu\rho} - \nabla_{\mu}R_{\nu\rho}, \;\; G_{\mu\nu} = -8\pi T_{\mu\nu} + \ldots .$$

**2-2.** Two categories of the general covariant field equations that are derived directly by linearization (and simultaneous parameterization), followed by first quantization (and applying the minimal coupling procedure) of the Lorentz invariant energy-momentum relation (defined for a particle with rest mass $m_0$), are as follows :

$$\breve{\nabla}_{\lambda}R_{\mu\nu\rho\sigma} + \breve{\nabla}_{\mu}R_{\nu\lambda\rho\sigma} + \breve{\nabla}_{\nu}R_{\lambda\mu\rho\sigma} = T^{\tau}{}_{\lambda\mu}R_{\tau\nu\rho\sigma} + T^{\tau}{}_{\mu\nu}R_{\tau\lambda\rho\sigma} + T^{\tau}{}_{\nu\lambda}R_{\tau\mu\rho\sigma} \,, \tag{1-1}$$

$$\breve{\nabla}_{\mu}R^{\mu}{}_{\nu\rho\sigma} - \frac{im_0^{(G)}}{\hbar} k_{\mu}R^{\mu}{}_{\nu\rho\sigma} = -J^{(G)}_{\nu\rho\sigma} \tag{1-2}$$

$$R_{\mu\nu\rho\sigma} = (\partial_{\nu}\Gamma_{\rho\sigma\mu} - \Gamma^{\lambda}{}_{\rho\nu}\Gamma_{\lambda\sigma\mu}) - (\partial_{\mu}\Gamma_{\rho\sigma\nu} - \Gamma^{\lambda}{}_{\rho\mu}\Gamma_{\lambda\sigma\nu}) \; ; \tag{1-3}$$

$$\breve{\nabla}_{\lambda} F_{\mu\nu} + \breve{\nabla}_{\mu} F_{\nu\lambda} + \breve{\nabla}_{\nu} F_{\lambda\mu} = 0, \tag{2-1}$$

$$\breve{\nabla}_{\mu} F^{\mu}{}_{\nu} = -J^{(E)}_{\nu}, \tag{2-2}$$

$$F_{\mu\nu} = \breve{\nabla}_{\nu} A_{\mu} - \breve{\nabla}_{\mu} A_{\nu} = (\nabla_{\nu} + \frac{im_0^{(EM)}}{2\hbar} k_{\nu}) A_{\mu} - (\nabla_{\mu} + \frac{im_0^{(EM)}}{2\hbar} k_{\mu}) A_{\nu}. \tag{2-3}$$

where $\Gamma^{\rho}{}_{\sigma\mu}$ is the affine connection, $i\hbar\breve{\nabla}_{\mu}$ is the energy-momentum quantum operator where $\breve{\nabla}_{\mu}$ is the general covariant derivative with torsion tensor $T_{\tau\mu\nu}$ that generates the invariant masses of the tensor fields $R_{\mu\nu\rho\sigma}$ and $F_{\mu\nu}$ ( $\breve{\nabla}_{\mu}$ also denotes formally the derivative operator without torsion, i.e. defined just by Christoffel symbols, however, it is not a covariant operator). In equations (1-1) – (2-3), $R_{\mu\nu\rho\sigma}$ and $F_{\mu\nu}$ denote the field strength tensors corresponding, respectively, to the Riemann curvature tensor (representing the gravitational field) and the Maxwell and Yang-Mills fields solely in (1+3) dimensional space-time. These fields in (1+2) space-time dimension, quantum mechanically, correspond to a new massive formalism of the bispinor fields of spin-1/2 and spin-3/2 particles [1].

The torsion tensor $T_{\tau\mu\nu}$ in the field equations (1-1) – (1-3) is given by:

$$T_{\tau\mu\nu} = \frac{im_0^{(G)}}{2\hbar}(g_{\tau\mu}k_{\nu} - g_{\tau\nu}k_{\mu}), \quad T_{\nu} = T^{\mu}{}_{\mu\nu} = (D-1)\frac{im_0^{(G)}}{2\hbar}k_{\nu} \tag{3}$$

\

and for the equations (2-1) – (2-3) the torsion is defined as follows:

$$T_{\tau\mu\nu} = \frac{im_0^{(EM)}}{2\hbar}(g_{\tau\mu}k_{\nu} - g_{\tau\nu}k_{\mu}), \quad T_{\nu} = T^{\mu}{}_{\mu\nu} = (D-1)\frac{im_0^{(EM)}}{2\hbar}k_{\nu} \tag{4}$$

where D is the number of space-time dimensions. Moreover, the source current tensors in the field equations (1-1) – (2-3) are defined necessarily by the following relations, respectively:

$$J^{(G)}_{\nu\rho\sigma} = -(\breve{\nabla}_\nu + \frac{im_0^{(G)}}{\hbar} k_\nu)\varphi^{(G)}_{\rho\sigma}, \quad J^{(E)}_\nu = -(\breve{\nabla}_\nu + \frac{im_0^{(EM)}}{\hbar} k_\nu)\varphi^{(EM)} \tag{5}$$

where $\varphi^{(G)}_{\rho\sigma}$ and $\varphi^{(EM)}$ are the given tensor quantities, $m_0^{(G)}$ and $m^{(EM)}$ are the invariant masses of the tensor fields $R_{\mu\nu\rho\sigma}$ and $F_{\mu\nu}$, $k_\mu = (c/\sqrt{g_{00}}, 0, ..., 0)$ is the general relativistic velocity of a static observer (that is a time-like covariant vector), and $A_\mu$ is the gauge potential vector.

Based on the torsion formalism of field equations (2-1) - (2-3) (corresponding to the Maxwell and Yang-Mills equations), and in agreement with the recent experimental (astronomical) data [9-12], the invariant mass of a new particle (identified as the massive gauge boson of U(1) symmetry group) has been calculated and specified as: ($m_0^{(E)} \approx 1.470696 \times 10^{-41}$ kg), where its invariant mass is generated by a coupling torsion tensor of the background space-time geometry.

It is noteworthy to note that our axiomatic approach could be considered in framework of theories which lie beyond the Standard Model [14, 15], as it also includes new consequences such as a certain massive formulation for the gravitational field.

In addition, we also should note that the field equations (1-1) – (2-3), quantum mechanically, would correspond to two different bispinor fields: they present a new massive form of bispinor fields of spin-1/2 and spin-3/2 particles, formulated solely in (1+2) dimensional space-time–where we necessary have $R_{20\rho\sigma} = R_{02\rho\sigma} = 0$ and $F_{20} = F_{02} = 0$ [6, 8]; These equations also present a new massive form of bispinor fields of spin-1 and spin-2 particles, formulated solely in (1+3) dimensional space-time [16, 17]. It should be emphasized here that the quantum mechanical solutions of these tensor equations are taken to be complex necessarily [6, 7, 8, 16, 17]. However, in the context of relativistic quantum mechanics, tensor equations (1-1) – (2-3) are also subject to a process of the $2^{nd}$ (canonical) quantization; then these equations would describe the bosonic fields in (1+3) dimensional space-time, and the fermionic fields in (1+2) dimensional space-time.

**2-3.** The tensor field equations (1-1) – (2-3), firstly and originally, are obtained in the following matrix formalism which are compatible with the matrix representation of Clifford algebras corresponding to various spin groups [18, 19]:

$$(i\hbar\alpha^{\mu}\breve{\nabla}_{\mu} - m_0\tilde{\alpha}^{\mu}k_{\mu})\Psi_R = 0, \tag{1-A}$$

$$(i\hbar\alpha^{\mu}\breve{\nabla}_{\mu} - m_0\tilde{\alpha}^{\mu}k_{\mu})\Psi_E = 0 \tag{2-A}$$

where
$$\alpha^{\mu} = \beta^{\mu} + \beta'^{\mu},\ \tilde{\alpha}^{\mu} = \beta^{\mu} - \beta'^{\mu} \tag{6}$$

$\Psi_R$, $\Psi_E$ are column matrices, $\beta^{\mu}$ and $\beta'^{\mu}$ are contravariant square matrices. These matrix quantities in (1+2) and (1+3) dimensional space-time are given as follows, respectively:

$$\beta^0 = \begin{bmatrix} 0 & 0 \\ 0 & -(\sigma^0+\sigma^1) \end{bmatrix},\ \beta'_0 = \begin{bmatrix} \sigma^0+\sigma^1 & 0 \\ 0 & 0 \end{bmatrix},\ \beta^1 = \begin{bmatrix} 0 & \sigma^2 \\ -\sigma^2 & 0 \end{bmatrix},\ \beta'_1 = \begin{bmatrix} 0 & \sigma^3 \\ -\sigma^3 & 0 \end{bmatrix},$$
$$\beta^2 = \begin{bmatrix} 0 & -\sigma^1 \\ -\sigma^0 & 0 \end{bmatrix},\ \beta'_2 = \begin{bmatrix} 0 & -\sigma^0 \\ -\sigma^1 & 0 \end{bmatrix};$$

$$\Psi_R = \begin{bmatrix} R_{10\rho\sigma} \\ 0 \\ R_{21\rho\sigma} \\ \varphi^{(R)}_{\rho\sigma} \end{bmatrix},\ \Psi_E = \begin{bmatrix} F_{10} \\ 0 \\ F_{21} \\ \varphi^{(EM)} \end{bmatrix},\ J^{(G)}_{\nu\rho\sigma} = -(\breve{\nabla}_{\nu} + \frac{im_0^{(G)}}{\hbar}k_{\nu})\varphi^{(G)}_{\rho\sigma},\ J^{(E)}_{\nu} = -(\breve{\nabla}_{\nu} + \frac{im_0^{(EM)}}{\hbar}k_{\nu})\varphi^{(EM)}$$

where $$\sigma^0=\begin{bmatrix}1&0\\0&0\end{bmatrix},\ \sigma^1=\begin{bmatrix}0&0\\0&1\end{bmatrix},\ \sigma^2=\begin{bmatrix}0&1\\0&0\end{bmatrix},\ \sigma^3=\begin{bmatrix}0&0\\-1&0\end{bmatrix} \quad \text{(7-1)}$$

In relations (7) it is assumed that $R_{20\rho\sigma}=R_{20\rho\sigma}=0,\ F_{20}=F_{02}=0$ ; in [1] we've shown that a certain discrete symmetric assumptions yield these conditions.

For (1+3) dimensional space-time we have:

$$\beta^0=\begin{bmatrix}0&0\\0&-(\gamma^0+\gamma^1)\end{bmatrix},\ \beta_0'=\begin{bmatrix}(\gamma^0+\gamma^1)&0\\0&0\end{bmatrix},\ \beta^1=\begin{bmatrix}0&\gamma^2\\-\gamma^3&0\end{bmatrix},\ \beta_1'=\begin{bmatrix}0&\gamma^3\\-\gamma^2&0\end{bmatrix},$$
$$\beta^2=\begin{bmatrix}0&\gamma^4\\\gamma^5&0\end{bmatrix},\ \beta_2'=\begin{bmatrix}0&-\gamma^5\\-\gamma^4&0\end{bmatrix},\ \beta^3=\begin{bmatrix}0&\gamma^6\\-\gamma^7&0\end{bmatrix},\ \beta_3'=\begin{bmatrix}0&\gamma^7\\-\gamma^6&0\end{bmatrix},$$

$$\Psi_R=\begin{bmatrix}R_{10\rho\sigma}\\R_{20\rho\sigma}\\R_{30\rho\sigma}\\0\\R_{23\rho\sigma}\\R_{31\rho\sigma}\\R_{12\rho\sigma}\\\varphi^{(G)}_{\rho\sigma}\end{bmatrix},\ \Psi_F=\begin{bmatrix}F_{10}\\F_{20}\\F_{30}\\0\\F_{23}\\F_{31}\\F_{12}\\\varphi^{(EM)}\end{bmatrix},\ J^{(G)}_{\nu\rho\sigma}=-(\breve{\nabla}_\nu+\frac{im_0^{(G)}}{\hbar}k_\nu)\varphi^{(G)}_{\rho\sigma},\ J^{(E)}_\nu=-(\breve{\nabla}_\nu+\frac{im_0^{(EM)}}{\hbar}k_\nu)\varphi^{(EM)} \quad \text{(8)}$$

where

$$\gamma^0=\begin{bmatrix}1&0&0&0\\0&1&0&0\\0&0&0&0\\0&0&0&0\end{bmatrix},\ \gamma^1=\begin{bmatrix}0&0&0&0\\0&0&0&0\\0&0&1&0\\0&0&0&1\end{bmatrix},\ \gamma^2=\begin{bmatrix}0&0&0&1\\0&0&0&0\\0&0&0&0\\-1&0&0&0\end{bmatrix},\ \gamma^3=\begin{bmatrix}0&0&0&0\\0&0&-1&0\\0&1&0&0\\0&0&0&0\end{bmatrix}.$$
$$\gamma^4=\begin{bmatrix}0&0&0&0\\0&0&0&1\\0&0&0&0\\0&-1&0&0\end{bmatrix},\ \gamma^5=\begin{bmatrix}0&0&-1&0\\0&0&0&0\\1&0&0&0\\0&0&0&0\end{bmatrix},\ \gamma^6=\begin{bmatrix}0&0&0&0\\0&0&0&0\\0&0&0&1\\0&0&-1&0\end{bmatrix},\ \gamma^7=\begin{bmatrix}0&-1&0&0\\1&0&0&0\\0&0&0&0\\0&0&0&0\end{bmatrix}. \quad \text{(8-1)}$$

It is noteworthy that the 4x4 matrices $\gamma^\mu$ in (8-1) fully generate the Lorentz Lie algebra in (1+3) dimensions as well.

**2-4**. In Ref. [1], we've also shown that from the massless case (*i.e.* $m_0^{(G)} = 0$) of the gravitational field equations (1-1) – (1-2) (or their equivalent matrix formulation i.e. equation (1-A)), i.e.,

$$\nabla_\lambda R_{\mu\nu\rho\sigma} + \nabla_\mu R_{\nu\lambda\rho\sigma} + \nabla_\nu R_{\lambda\mu\rho\sigma} = 0 \,, \tag{9-1}$$

$$\nabla_\mu R^\mu{}_{\nu\rho\sigma} = -J^{(G)}_{\nu\rho\sigma} \tag{9-2}$$

the Einstein field equations are obtained directly as follows. First, using the relation $R_{\sigma\mu} = -R^\nu_{\sigma\mu\nu}$ (as definition of the Ricci tensor) by contraction of the 2nd Bianchi identity (9-1) we may get:

$$\nabla_\sigma R_{\mu\nu\rho}{}^\sigma = \nabla_\nu R_{\mu\rho} - \nabla_\mu R_{\nu\rho} \tag{10}$$

(However, in our approach the Ricci tensor $R_{\mu\nu}$ is defined, basically, by relation (10)).

Then from (9), (10) and the following definition for current $J^{(G)}_{\nu\rho\sigma}$,

$$J^{(G)}_{\nu\rho\sigma} = -8\pi(\nabla_\sigma T_{\nu\rho} - \nabla_\rho T_{\nu\sigma}) + 8\pi B(\nabla_\sigma T g_{\nu\rho} - \nabla_\rho T g_{\nu\sigma}) \,, \tag{11}$$

where $T_{\mu\nu}$ is the stress-energy tensor (and $T = T^\mu{}_\mu$), $g_{\mu\nu}$ is the metric and $B$ is a constant (which would be specified for each space-time dimension), we obtain the Einstein field equations:

$$R_{\mu\nu} = -8\pi(T_{\mu\nu} - BTg_{\mu\nu}) - \Lambda g_{\mu\nu} \tag{12}$$

where $\Lambda$ is a cosmological constant (emerged naturally in the course of obtaining (12)). For (1+1) dimensional case from (12) we get: (however, matrix formalism of (12) in 1+1 dimensions defined by (1-A), doesn't hold one of the basic assumed discrete symmetries for matrix equations (1-A) and (2-A))

$$R_{\mu\nu} = -4\pi T g_{\mu\nu} + \tfrac{1}{2}\Lambda g_{\mu\nu} \tag{13}$$

where $B = 0$. For (1+2) dimensions we have

$$R_{\mu\nu} - \tfrac{1}{2} R g_{\mu\nu} = -8\pi T_{\mu\nu} - 2\Lambda g_{\mu\nu} \tag{14}$$

where $B = 1$. For (1+3) dimensional space-time, we obtain

$$R_{\mu\nu} - \tfrac{1}{2} R g_{\mu\nu} = -8\pi T_{\mu\nu} - \Lambda g_{\mu\nu} \tag{15}$$

where $B = 1/2$.

**2-5.** Based on the definite mathematical formalism of this axiomatic approach, along with the C, P and T symmetries (represented basically by the corresponding quantum operators) of the fundamentally derived field equations, in Ref. [1], it has been concluded that the universe could be realized solely with the (1+2) and (1+3)-dimensional space-times (where this conclusion, in particular, is based on the T-symmetry of these equations). In addition, on the basis of these discrete symmetries of derived field equations, it has been also shown that only left-handed particle fields (along with their complementary right-handed fields) could be coupled to the corresponding (any) source currents. Furthermore, it has been shown that the metric of background space-time is diagonalized for the uniquely derived fermion field equations (defined and expressed solely in (1+2)-dimensional space-time), where this property generates a certain set of additional symmetries corresponding uniquely to the "SU(2)_L × U(2)_R" symmetry group for spin-1/2 fermion fields (representing "1+3" generations of four fermions, including a group of eight leptons and a group of eight quarks), and also the "SU(2)_L × U(2)_R" and SU(3) gauge symmetry groups for spin-1 boson fields coupled to the spin-1/2 fermionic source currents. Hence, along with the known elementary particles, eight new elementary particles, including four new charge-less right-handed spin-1/2 fermions (two leptons and two quarks, that could be represented by "z_e , z_n ; z_u, z_d", where two quarks "z_u , z_d" in particular are mixed solely with the triple compositions of antiquarks in anti-baryons' structures), a spin-3/2 fermion, and also three new spin-1 massive bosons (where in particular, the new boson is complementary right-handed particle of ordinary boson), are predicted uniquely by this axiomatic approach.

**2-6.** According to the unique mathematical structure of the derived field equations (2-1) – (2-3) corresponding to the Maxwell's (and the Yang-Mills) equations, in Ref. [1] we've shown that magnetic monopoles (in contrast with electric monopoles) could not exist in nature.

## Acknowledgment :

Special thanks are extended to Prof. and Academician Vitaly L.Ginzburg (Russia, Nobel Laureate), Prof. and Academician Dmitry V.Shirkov (Russia), Prof. Leonid A. Shelepin (Russia), Prof. VladimirYa. Fainberg (Russia), Prof. Wolfgang Rindler (USA), Prof. Roman W. Jackiw (USA), Prof. Sir Roger Penrose (UK), Prof. Koji Nakatogawa (Japan), Prof. Steven

Weinberg (USA, Nobel Laureate), Prof. Ezra T.Newman (USA), Prof. Graham Jameson (UK), Prof. Sergey A. Reshetnjak (Russia), Prof. Sir Michael Atiyah (UK) (who, in particular, kindly encouraged me to continue this work as a new unorthodox mathematical approach to fundamental physics), and many others for their support and valuable guidance during my studies and research.